\documentclass[fleqn,usenatbib]{mnras}

\usepackage{newtxtext,newtxmath}
\usepackage{layouts}

\usepackage[T1]{fontenc}

\DeclareRobustCommand{\VAN}[3]{#2}
\let\VANthebibliography\thebibliography
\def\thebibliography{\DeclareRobustCommand{\VAN}[3]{##3}\VANthebibliography}

\usepackage{graphicx}	
\usepackage{amsmath}	

\usepackage{tablefootnote}
\usepackage[flushleft]{threeparttable} 

\newcommand{\msun}{\, M_\odot}
\title[Limits on Dark Matter Compact Objects]{Limits on Dark Matter Compact Objects implied by Supermagnified Stars in Lensing Clusters}

\author[Claudi Vall M\"uller et al.]{
Claudi Vall M\"uller$^{1}$,
Jordi Miralda-Escudé$^{1,2,3}$\thanks{miralda@icc.ub.edu}.
\\
$^{1}$Institut de Ci\`encies del Cosmos, Universitat de Barcelona, Barcelona, Spain \\
$^{2}$Instituci\'o Catalana de Recerca i Estudis Avan\c cats, Barcelona, Spain \\
$^{3}$Institut d'Estudis Espacials de Catalunya, Barcelona, Spain \\
}

\date{Accepted XXX. Received YYY; in original form ZZZ}

\pubyear{2024}

\begin{document}
\label{firstpage}
\pagerange{\pageref{firstpage}--\pageref{lastpage}}
\maketitle

\begin{abstract}
Supermagnified stars are gravitationally lensed individual stars that are located close to a caustic of a lensing galaxy cluster, and have their flux magnified by a large enough factor (typically $\sim$ 1000) to make them detectable with present telescopes.  The maximum magnification is limited by microlensing caused by
intracluster stars or other compact objects, which create a network of corrugated critical lines with an angular width proportional to the surface density of microlenses. We consider a set of 9 cases of supermagnified stars reported in the literature, and derive an upper limit on the surface density of compact objects, such as
primordial black holes, that might be present as a fraction of the dark matter in addition to known intracluster stars. Any such additional compact objects would widen the corrugated critical line network and therefore the width of the distribution of supermagnified stars around the modeled critical lines of the lens. We find that any compact objects, including primordial black holes, with masses above $\sim 10^{-6} \msun$ (below which the microcaustics are closer together than the typical angular size of supermagnified stars) cannot account for more than $\sim$ 2\% of the dark matter.
\end{abstract}



\section{Introduction}

Several independent observations have demonstrated beyond reasonable doubt that the known baryonic matter accounts for only $\sim$ 15\% of the matter in the Universe, and the remaining 85\% is a form of collisionless matter, designated as dark matter since the early work of \cite{Zwicky33} on galaxy clusters. The first clear evidence for dark matter initially came from galaxy rotation curves and galaxy velocities within groups \citep{OPY74,EKS74,Rubin78}, but present evidence has confirmed the existence of cold dark matter \citep[meaning collisionless matter with no appreciable initial velocity dispersion; see][]{P82} from a wide variety of other independent observations: the Cosmic Microwave Background fluctuations, which can be explained only with this ratio of baryonic to dark matter \citep{Planck20,Mad23,T24}; lensing and X-ray observations of clusters that agree with the same baryon fraction and with the predicted density profiles \citep[e.g.,][]{Umetsu18}; the large-scale structure galaxy correlations \citep{EBOSS21}; and the agreement with the baryon density inferred from primordial nucleosynthesis \citep{GF22}.

Dark matter may be in the form of compact objects, which would be detectable through gravitational lensing. The baryon density derived from primordial nucleosynthesis and CMB observations rules out compact objects made of baryonic matter, but primordial black holes (PBH) formed in the early universe from the collapse of the hot plasma in regions of large amplitude primordial perturbations can account for cold dark matter without requiring the addition of any particles to the Standard Model of particle physics \citep{Hawking71}. Although a number of astrophysical observations have constrained the contribution of PBH to the dark matter of the Universe \citep[see][for a review]{Carr21}, the asteroid-mass window $10^{-16} \msun < m_b < 10^{-10} \msun$ remains completely open. Higher mass PBH are ruled out as constituents of all the dark matter by microlensing surveys in the Magellanic Clouds, although a small fraction of  $\sim$ 10\% of dark matter in compact objects is still allowed \citep{MACHO98,MACHO01,EROS07,OGLE3-11a,OGLE3-11b,Griest14,Smyth20}. This fraction is even lower in the stellar mass range from the most recent results of the OGLE experiment \citep{Mroz24,MrozN24}.

Microlensing observations can provide independent constraints on the fraction of dark matter in compact objects from a different type of observations: highly magnified luminous stars lensed by clusters of galaxies that lie very near a lensing caustic. The high magnification allows a luminous star at a cosmological distance to be observed via our most powerful telescopes \citep{Miralda91}. The presence of compact objects near the lensing critical line of the cluster replaces the smooth critical line with a corrugated network of micro-critical lines, and the higher the surface density of compact objects, the broader this network over which magnified images reaching the highest magnifications may be observed \citep{VDM17,Oguri18}. The discovery of such highly magnified stars \citep{KDR18} has opened the door to use these observations to constrain the presence of any compact objects in the dark matter of a lensing cluster (plus any dark matter along the line of sight contributing to the lensing deflection), which is over the contribution from intracluster stars that are known to be present.

In this paper we analyze the most interesting cases of highly magnified stars that have been reported in the literature to derive an upper limit to the fraction of dark matter in lensing clusters that may be in the form of dark, compact objects. Section \ref{sec:distribution} reviews the properties of the network of micro-critical lines that are relevant to derive this upper limit, and introduces an analytical approximation for the density of supermagnified images around this network. Our method to derive an upper limit on compact objects is described in Section \ref{sec:widths}, together with the list of supermagnified stars we use in this work. Our results are presented in Section \ref{sec:results}, with a discussion on the dependence on the lens mass in Section \ref{sec:advanced}, and we summarize our conclusion in Section \ref{sec:conclusions}.

\section{The distribution of highly magnified images around the smooth critical line}
\label{sec:distribution}

We now discuss the expected distribution of highly magnified images in the observed image plane of a lens.

In general, highly magnified images are located in the vicinity of the critical lines, defined as the curves in the image plane where the determinant of the magnification matrix is zero. If the lens surface density is a differentiable function, without structure at scales less than some characteristic smoothing scale, the critical lines are also smooth and differentiable curves. A real lensing cluster, however, has some fraction of its mass in compact objects: at the very least, intracluster stars acting as point masses are present. It is then convenient to define the macro-model of the lens as a smoothed version of the true lens, with a surface density resulting from smoothing the point masses over a scale that is much smaller than the total lens deflection, but much larger than the separation among neighboring microlenses.

We follow \cite{VDM17}, hereafter V17, to establish our nomenclature and notation. The macro-critical line (MaCL) is defined as the critical line of the lens macro-model, including the smoothed contribution from point masses, which we refer to as microlenses hereafter. The smooth critical line (SCL), on the other hand, is the critical line of all the rest of the lens mass, without including the microlenses, which is assumed to be intrinsically smooth on small scales. The true critical line takes into account the microlenses (the known intracluster stars, plus any other hypothetical compact objects that may account for some fraction of the dark matter), and is called the micro-critical line (MiCL). The MiCL is highly complex, and forms a corrugated band of microcaustics that meander around many microlenses. As discussed in V17, this corrugated band is centered on the SCL and not the MaCL, with an angular width
given by (see eq.\ 22 in V17)
\begin{equation}\label{eq:rw}
    r_w= \frac{\kappa_\star}{g} ~,
\end{equation}
where $\kappa_\star$ is the contribution to the lensing convergence from the smoothed point masses, and $g$ is the angular gradient modulus of the eigenvalue of the macro-model magnification matrix that vanishes at the MaCL (note that $g$ is written as $d$ in V17). Defining $s$ to be the angular separation from the SCL in the lens plane, with a positive sign for points outside the SCL and negative sign inside,
the MiCLs are typically connected to each other to form an intricate, corrugated network in the region $|s| <r_w$, which is affected everywhere by many point masses. Outside the corrugated network, at $|s| >r_w$, the MiCLs are usually disconnected and form closed curves around each point mass, with the shape of the infinite symbol in the approximation of a single point mass with fixed external shear \citep{Chang79}. The total width of the corrugated band is therefore $2r_w$. Because the macro-model has a greater surface density by $\kappa_\star$ than the smooth component, the MaCL is external to the SCL and is shifted outwards by an angular separation $r_w$ from the SCL, just at the outer edge of the corrugated band.

We first review the analytical derivation for how the maximum magnification reached in MiCL crossings depends on $s/r_w$. The maximum magnification $\mu_p$ that is reached when a source star of angular radius $\theta_s$ crosses a MiCL depends on the local eigenvalue gradient $g_\star$, computed including the point masses with no smoothing (whereas $g$ is the eigenvalue gradient of the macro-lens model). We assume for simplicity that all microlenses have the same mass, corresponding to an Einstein radius $\theta_\star$ (we discuss in Section \ref{sec:advanced} the likely effect of a distribution of microlens masses).
The peak magnification is
$\mu_p\propto (g_\star\theta_s\sin\alpha_\star)^{-1/2}$ (eq.\ 27 in V17),
where $\alpha_\star$ is the angle formed by the MiCL and the principal axis of the eigenvalue that cancels on the MiCL.
The average separation between neighboring point masses is $\theta_\star/\kappa_\star^{1/2}$. Within the corrugated band, the MiCL is a curve meandering around point masses at this typical separation, where the typical shear from point masses is $\lambda_\star\sim \kappa_\star$ and its gradient is $g_\star\sim \kappa_\star^{3/2}/\theta_\star$. Therefore, the distribution of peak magnifications of MiCL crossings is roughly constant within the width $r_w$, with the typical value scaling as 
\begin{equation}
\mu_p\propto (\theta_\star/\theta_s)^{1/2} \kappa_\star^{-3/4} \qquad (|s| < r_w)~.
\end{equation}
We assume that the distribution of the angle $\alpha_\star$ is random and independent of $r$, so it does not affect the proportionality relations discussed here.

However, when moving outside the network at a separation $|s| >r_w$ from the SCL, MiCLs are closed curves around individual point masses of a characteristic size $\theta_i\sim \theta_\star(\kappa_\star |s|/r_w)^{-1/2}$, where the characteristic shear is now $\lambda_\star \sim \kappa_\star (|s|/r_w)$ and the eigenvalue gradient is typically $g_\star\sim \lambda_\star/\theta_i \sim (\kappa_\star\, |s|/r_w)^{3/2}/\theta_\star$. The reason is that the shear of a point mass drops with the angular separation as $\theta_i^{-2}$, and the shear from microlenses that needs to be added to the macro-model to reach a zero of the magnification eigenvalue increases as $s/r_w$. The peak magnification that is reached is therefore reduced to
\begin{equation}
\mu_p\propto (\theta_\star/\theta_s)^{1/2} (\kappa_\star\, |s|/r_w)^{-3/4} \qquad (|s| > r_w) ~.
\end{equation}
When expressed as a function of the source separation from the caustic of the smooth model, $y\propto s^2$, the peak magnification drops as $\mu_p\propto y^{-3/8}$ outside the corrugated band of micro-caustics, and is flat inside (see eq.\ 27 in V17). 

Nevertheless, what we actually want to infer here is not the dependence of peak magnifications on $s$,
but the distribution of highly magnified images around the SCL that reach
a magnification above some threshold value, $\mu > \mu_t$. For this purpose,
we compute first the fraction of the image plane area where the magnification is above $\mu_t$, in the limit of high $\mu_t$ when this fraction is small, as
a function of $s$.
The region around each
MiCL that has magnification $\mu > \mu_t$ has a width proportional to $g_\star^{-1}$, assuming the source star is small enough to be fully inside this width from the microcaustic in the source plane.
Within $|s| <r_w$ this fraction is roughly constant because the distribution of $g_\star$ is also roughly independent of $s$. Outside the corrugated band,
the typical value of $g_\star$ on MiCLs grows as $(|s|/r_w)^{3/2}$, and the length of MiCLs is proportional to their typical size around each point mass, or to $(|s|/r_w)^{-1/2}$. Therefore, the fraction of the area with $\mu > \mu_t$, scales as 
\begin{equation}
{\rm Area} \sim g_\star^{-1}(|s|/r_w)^{-1/2}\sim (|s|/r_w)^{-2} ~.
\end{equation}

Because we have computed this area fraction at fixed $\mu$, the density of micro-images above $\mu_t$ from sources that are randomly distributed in the source plane must also scale in the same way: roughly constant in $|s|<r_w$ and falling as $(|s|/r_w)^{-2}$ at $|s|>r_w$.
Based on these approximations, we adopt the following model
for computing likelihood functions of the observed positions of highly magnified images:
\begin{equation}\label{eq:probi}
  p_i(s)\, ds = \frac{1}{\pi}\, \frac{ds/r_w}{ 1+ (s/r_w)^2} ~.
\end{equation}
We note that this expression for the function $p_i(s)$ assumes that the source star is small enough not to have reached the maximum possible magnification, which is produced when the stellar disk lies on the micro-caustic. Once this maximum magnification is reached, the probability density $p_i$ drops much faster with $s$ than in equation (\ref{eq:probi}), because most MiCL become too small to reach the required magnification.

Our adoption of equation (\ref{eq:probi}) for the functional form of the probability distribution of magnified images is done for simplicity in this brief and necessarily preliminary study. A more thorough analysis will naturally benefit from a detailed functional form derived from numerical simulations of microlensing.

Observations usually provide good constraints on the position of the MaCL of lensing models, when pairs of images of the same extended object are observed on large scales compared to the microlenses, which are at equal separations $r$ from the MaCL. As discussed earlier, the separation from the SCL is $s=r+r_w$, where $s$ and $r$ are defined to be positive outside the SCL and MaCL respectively, and negative inside. Therefore, our model for the distribution of magnified images expressed in terms of the separation from the MaCL is
\begin{equation}\label{eq:probir}
  p_i(r)\, dr = \frac{1}{\pi}\, \frac{dr/r_w}{ 1+ (1+r/r_w)^2} ~.
\end{equation}

\section{Corrugated band widths}
\label{sec:widths}
Our goal in this section is to measure the distribution of separations from the MaCL in a sample of supermagnified stars, to derive an upper limit on the surface density of microlenses.
For this purpose, we select a list of 12 supermagnified star candidates that had been reported as this work was being done, listed in Table \ref{tab:parameters_table}.
The list starts with the initial discovery of Icarus (\cite{KDR18}), up to the recent report of Mothra (\cite{DSY23}).

We consider these events are most likely to be true supermagnified stars. Supermagnified stars should follow the distribution approximated by equation (\ref{eq:probir}) only when their detection requires a minimum magnification threshold to make them visible, and the star size is smaller than the separation from the microcaustic when magnified.
Single supermagnified star candidates may be confused with other objects, mainly compact globular clusters that remain unresolved, but this is unlikely for most of these candidates because clusters that are compact enough are rare \citep[see][]{WCD22}. The main criteria to consider these events as good supermagnified star candidates are photometric data that are consistent with the spectra of some of the most luminous stars (generally supergiants or the most massive main-sequence stars), and variability on timescales of $\sim$ days that is consistent with being caused by MiCL crossings of sources with radii typical of these most luminous stars, with a lack of observed counterimages which are typically not seen owing to insufficient magnification.

The candidates in Table \ref{tab:parameters_table}
generally satisfy these criteria. There are doubts, however, on the nature of the two Spock objects and Godzilla (events labeled as 2, 3 and 6 in Table \ref{tab:parameters_table}), which may be more consistent with flaring luminous blue variables (similar to Eta Carinae) or recurrent novae \citep{RBB18,DPK22}, especially in view of their high brightness. In addition, for these cases it is difficult to determine precisely where the MaCL lies \citep[see in particular the discussion in][]{DPK22}, and therefore the separation $r$ from the observed variable images to the MaCL. The position of the MaCL is best determined when pairs of images of surface brightness irregularities in the source galaxy can be identified, which indicate that the MaCL lies precisely in between the image pair (assuming the pair separation is much less than the MaCL size, so that a linear variation of the magnification matrix from the MaCL is a good approximation). This generally gives a high confidence in the modelled separations $r$ listed in column 9 of Table \ref{tab:parameters_table} for the remaining 9 cases,
all of which have consistent photometry and variability to be microlensing events of single stars or compact systems of few stars (binaries or multiple hierarchical binaries).

From the discussion in the previous section, we note that the separation from the SCL is obtained from the separation from the MaCL as $s=r+r_w=r+\kappa_\star/g$, and that the sign of $r$ listed in Table \ref{tab:parameters_table} is positive outside the MaCL. The corrugated band where the magnifications are highest and the largest density of supermagnified stars are expected is centered on the SCL, at $s=0$.

For each of the events, we compute the convergence on the MaCL $\kappa_0$, the eigenvalue inverse gradient $g^{-1}$, and the angular separation from the image to the MaCL $r$, using publicly available galaxy cluster lens models, such as the Hubble Frontier Fields (HFF) (\cite{LKC17-HFF}), CLASH (\cite{Postman12CLASH}) or RELICS (\cite{Coe19RELICS}) programs. For each model we use maps of the magnification matrix parameters, with the ratio of angular diameter distances $D_{\rm LS}/D_{\rm S}$ rescaled to the source redshift to infer $\kappa_0$ and $g^{-1}$. We evaluate the eigenvalue gradient at the MaCL, $\mathbf{g} = - (\mathbf{\nabla} \kappa)_0 - (\mathbf{\nabla} \gamma)_0$, at the point of the model MaCL which is connected to the star image by a segment that is parallel to the principal axis of the vanishing eigenvalue (i.e., the direction of image elongation). 
The derivatives of the convergence and shear along this segment are evaluated by subtraction of the values at points separated by $\sim 10$ times the pixel size in each lens model sky map. The presented values of $\kappa_0$ and $g^{-1}$ are obtained by averaging over the models listed in Table \ref{tab:parameters_table}. Statistical dispersions are generally small ($\sim 2\%$ for $\kappa_0$ and $\sim 10\%$ for $g^{-1}$), with exception of cases 1 and 5, where they reach values of $\sim 10\%$ and $\sim 40\%$ for $\kappa_0$ and $g^{-1}$, respectively.

We list also the convergence contributed by stars that is implied by the intracluster light brightness near the image, $\kappa_\star$, that is reported in the listed references for 5 of the 9 considered events. Here we simply list the surface density values directly as reported by the authors in each reference where it is given, although surface densities are obviously inferred from the observed surface brightness, generally assuming a Chabrier initial mass function, or a Salpeter one with a calibrated mass lower limit. This should be improved in the future using the same standardized initial mass function for all cases.
We then compute $\kappa_\star = \Sigma_\star/\Sigma_\text{crit}$. This measurement of $\Sigma_\star$ is not reported yet for the other 4 cases.

\begin{figure}
\centering
\includegraphics[width=0.95\linewidth]{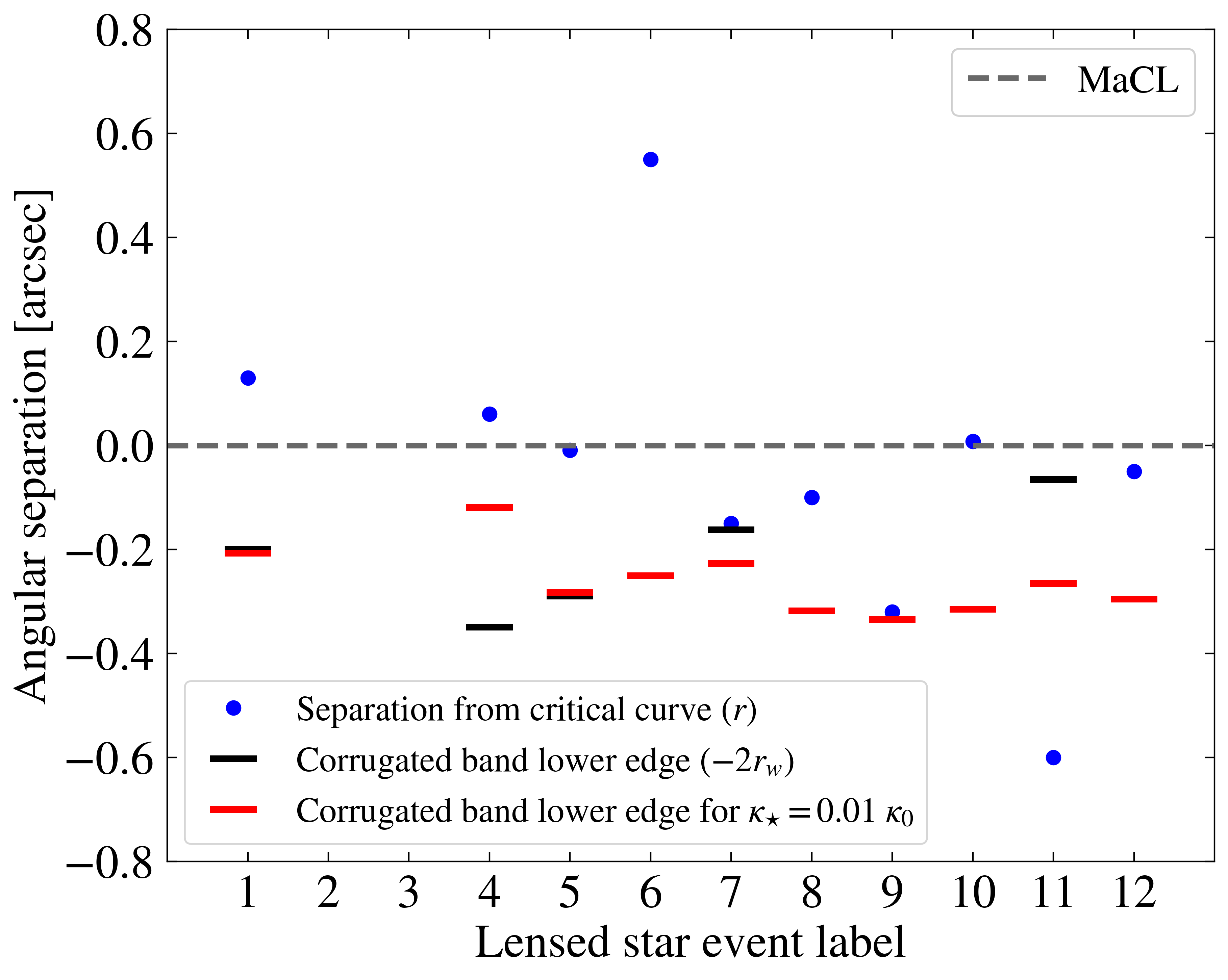}
\caption{Angular separation $r$ (blue dots) from each of the 12 highly magnified star images to the model macro critical line (MaCL, grey dashed line, which is the outer edge of the corrugated band at $r=0$), compared to the inner edge of the corrugated band of micro-critical lines (black bars) when the surface density $\kappa_\star$ is that of the reported intracluster light, and when it is $\kappa_\star = 0.01 \ \kappa_0$ (red bars). The corrugated band is centered at $r=-r_w$ (the SCL), while its inner edge shown by the bars is at $r=-2r_w$.
Events 2, 3 and 6 are excluded from our analysis because of uncertainties on the nature of the sources and the lens model of the MaCL.}
\label{fig:critical curve distances plot}
\end{figure}

The predicted corrugated band half widths, $r_w$, are then computed using equation (\ref{eq:rw}), and are indicated at the position $r=-2r_w$ as black horizontal bars in Figure \ref{fig:critical curve distances plot}, for the 5 events with a known value of $\kappa_\star$. 
In this way, the dashed line at $r=0$ ($s=+r_w$) indicates the outer edge of the corrugated band, and the bars at $s=r+r_w=-r_w$ indicate the inner edge, while the band center is in between at $r=-r_w$.
We also show as red horizontal bars the bandwith corresponding to a total surface density of compact objects equal to 1\% of the total mass surface density of the lensing model (i.e., $\kappa_\star = 0.01 \kappa_0$), again at $r=s-r_w=-2r_w$. The observed angular separations between the star images and each model MaCL are shown as blue dots. For event number 6 we did not have the magnification matrix of the lens model available and the red bar indicates the mean of the other lenses.

We see that generally, the images are observed at locations compatible with the expected center and width of the corrugated band. For the distribution assumed in equation (\ref{eq:probir}), we expect half of the images to lie between the dashed line at $r=0$ and the black or red bars depending on our assumed $\kappa_\star$. As mentioned in Section \ref{sec:distribution}, the nature of event 6 is questionable, and the only other event that is a substantial outlier is event 11, which would lie very far from the corrugated band if we believe the very small value of $\kappa_\star$ contributed by intracluster stars given by \cite{MCZ23}. Event 11 may be a true outlier in the distribution of highly magnified stars or may have its stellar surface density underestimated.

From this list of cases, it is not yet possible to check if most magnified stars are inside the MaCL, at $r<0$, as expected, but this should become clear as more cases of supermagnified stars with good models and well measured $\kappa_\star$ become available. If we take the red bars as an example, the assumption of $\kappa_\star =0.01 \kappa_0$ results in 5 out of 9 cases within the band, with 3 outside the outer edge and 1 inside the inner edge, compared to the prediction of 50\% of images in the corrugated band from the simple model of equation (\ref{eq:probir}).

\section{Results}
\label{sec:results}

\begin{table*}
    \begin{threeparttable}
    
    \begin{tabular}{cccccccccc}
    \hline
    Event label & Lens cluster & Star name & $z_L$ & $z_S$ & $\kappa_0$ & $g^{-1}$ & $\kappa_\star$ & $r$ & AB mag \\
     &  &  &  &  &  & [arcsec] &  & [arcsec] & [mag] \\
    \hline
    1 \tnote{a} & MACS J1149 & Icarus \citep{KDR18} & 0.54 & 1.49 & 0.832 & 12.46 & 0.0080 & $+$ 0.13 & 25-26 (F125W) \\
    2 \tnote{b} & MACS J0416 & Spock-NW \citep{RBB18} & 0.397 & 1.0054 & 0.742 & 10.29 & 0.0028 & ... & 25-26.5 (F814W) \\
    3 \tnote{b} & MACS J0416 & Spock-SE \citep{RBB18} & 0.397 & 1.0054 & 0.629 & 10.29 & 0.0028 & ... & 23-26.5 (F160W) \\
    4 \tnote{c} & MACS J0416 & Warhol \citep{KDV19} & 0.397 & 0.94 & 0.686 & 8.74 & 0.02 & $+$ 0.06 & 26.25 (F125W) \\
    5 \tnote{d} & WHL0137–08 & Earendel \citep{WCD22} & 0.566 & 6.2 & 0.590 & 24.01 & 0.0060 & $-$ 0.009 & 27.2 (F150W) \\
    6 \tnote{e} & PSZ1 G311.65 & Godzilla \citep{DPK22} & 0.443 & 2.37 & ... & ... & ... & $+$ 0.55 & $\approx$ 22 (F814W) \\
    7 \tnote{f} & Abell 2744 & LS7 \citep{CKT22} & 0.308 & 2.65 & 0.726 & 15.68 & 0.0052 & $-$ 0.15 & 27.05 (F150W) \\
    8 \tnote{g} & MACS J0647 & star-1 \citep{MZJ23} & 0.591 & 4.8 & 0.544 & 29.28 & ... & $-$ 0.1 & 27.343 (F227W) \\
    9 \tnote{g} & MACS J0647 & star-2 \citep{MZJ23} & 0.591 & 4.8 & 0.549 & 30.59 & ... & $-$ 0.32 & 28.330 (F227W) \\
    10 \tnote{h} & El Gordo & Quyllur \citep{DMA23} & 0.870 & 2.1878 & 0.662 & 23.84 & ... & $+$ 7.5 $\cdot 10^{-3}$  & 25.5 (F356W) \\
    11 \tnote{i} & Abell 370 & LS11 \citep{MCZ23} & 0.375 & 1.2567 & 0.703 & 18.89 & 0.0017 & $-$ 0.6 & 29.51 (F200LP) \\
    12 \tnote{j} & MACS J0416 & Mothra \citep{DSY23} & 0.397 & 2.091 & 0.936 & 15.78& ... & $-$ 0.05 & $\approx$ 27.8 (F200W) \\
        \hline
    \end{tabular}
       \caption{List of 12 supermagnified star candidates, with lens and source redshifts $z_L$ and $z_S$, convergence at image position $\kappa_0$, eigenvalue inverse gradient $\mathbf{g} = \mathbf{\nabla}(-\kappa -\gamma$) at the macro critical curve (MaCL), angular separation from source to MaCL $r$ (with the positive direction pointing towards the outer side), and observed AB magnitude of each transient at given filter (AB mag). The values of $\kappa_0$ and $g^{-1}$ are the average of the set of publicly available lens models for each lens.}
    \label{tab:parameters_table}

    \begin{tablenotes}
    \item[a] CATS (v4.1) (\cite{Richard14LS1Cats4.1}), Zitrin-ltm (v1) (\cite{Zitrin09LS1Zltm1}; \cite{Zitrin13LS1Zltm1}), GLAFIC (v3) (\cite{Oguri10LS1GLAFIC3};\cite{Kawamata16LS1GLAFIC3}) and Keeton (v4) (\cite{McCully14LS1Keeton4}, \cite{Ammons14LS1Keeton4}, \cite{Keeton10LS1Keeton4}).
    \item[b] Williams (v4) (\cite{Sebesta16LS2Williams4}, \cite{Liesenborgs06LS2Williams4}) and Bradač (v3) (\cite{Hoag16LS2Bradac3}, \cite{Bradac05LS2Bradac3}, \cite{Bradac09LS2Bradac3}).
    \item[c] Caminha (v4) (\cite{Caminha17LS4Caminha4}), GLAFIC (v4) (\cite{Oguri10LS1GLAFIC3}, \cite{Kawamata16LS1GLAFIC3}, \cite{Kawamata18LS4Glafic4}), Keeton (v4) (\cite{McCully14LS1Keeton4}, \cite{Ammons14LS1Keeton4}, \cite{Keeton10LS1Keeton4}), Sharon (v4 Cor.) (\cite{Johnson14LS4Sharon4.2}, \cite{Jullo07LS4Sharon4.2}), Zitrin-ltm-gauss (v3) and Zitrin-nfw (v3) (\cite{Zitrin09LS1Zltm1}, \cite{Zitrin13LS1Zltm1}).
    \item[d] Zitrin-ltm (v1) (\cite{Zitrin09LS1Zltm1}, \cite{Zitrin15LS5Zltm1}), GLAFIC (v1) (\cite{Oguri10LS1GLAFIC3}), WSLAP (v1) (\cite{Sendra14LS5WSLAP1}) and Lenstool (v1) (\cite{Johnson14LS4Sharon4.2}, \cite{Jullo07LS4Sharon4.2}).
    \item[e] No public lens models found.
    \item[f] Sharon (v4 Cor.) (\cite{Johnson14LS4Sharon4.2}, \cite{Jullo07LS4Sharon4.2}).
    \item[g] Zitrin-ltm-gauss (v2) and Zitrin-nfw (v2) (\cite{Zitrin09LS1Zltm1}, \cite{Zitrin13LS1Zltm1}).
    \item[h] Diego (v1) (\cite{DMA23}).
    \item[i] GLAFIC (v4) (\cite{Oguri10LS1GLAFIC3}, \cite{Kawamata16LS1GLAFIC3}, \cite{Kawamata18LS4Glafic4}), Williams (v4) (\cite{Sebesta19LS11Williams1}) and Keeton (v4) (\cite{McCully14LS1Keeton4}, \cite{Ammons14LS1Keeton4}, \cite{Keeton10LS1Keeton4}).
    \item[j] Zitrin-ltm-gauss (v3) (\cite{Zitrin09LS1Zltm1}, \cite{Zitrin13LS1Zltm1}).
    \end{tablenotes}
    \end{threeparttable}

\end{table*}

We now proceed to compute a maximum likelihood estimate of the optimal value of the total surface density of microlenses to explain the distribution of separations in Figure \ref{fig:critical curve distances plot}, and an upper limit to the fraction of dark matter that may be composed of compact objects. Initially, we assume for simplicity that any compact objects accounting for part of the dark matter have a similar mass distribution as the intracluster stars, and in the next section we discuss the range of masses that our limit actually applies to.

Using the probability density function $p_i(r)$ of equation (\ref{eq:probir}), we define the likelihood function   
\begin{equation}
\ln \mathcal{L} (\kappa_\text{b}) = \sum_i \ln p_i (\kappa_\text{b}) ~,
\end{equation}
where $p_i(r_i, g_i, \kappa_\text{b})$ is the probability density of observing a stellar microlensing event separated by $r_i$ from the MaCL of event $i$, given a microlensing band of width $r_{w, i} = (\kappa_{\star, i} + \kappa_\text{b})/g_i$. We assume that the total contribution from point masses to microlensing is the sum of the known intracluster light ($\kappa_\star$) and a hypothetical contribution from dark matter compact objects, such as primordial black holes ($\kappa_\text{b}$). We then find the optimal value of $\kappa_\text{b}$ that maximizes this likelihood function. We use all 9 events considered reliable. The value of $\kappa_\star$ is measured only for 5 of these events, and for the other 4 we simply use the average value of $\kappa_\star$ of all 7 events where it is measured (including events 2 and 3 in Table \ref{tab:parameters_table} that are not considered reliable as supermagnified stars), which is $\bar\kappa_\star= 6\times 10^{-3}$.

\begin{figure}
\centering
\includegraphics[width=0.95\linewidth]{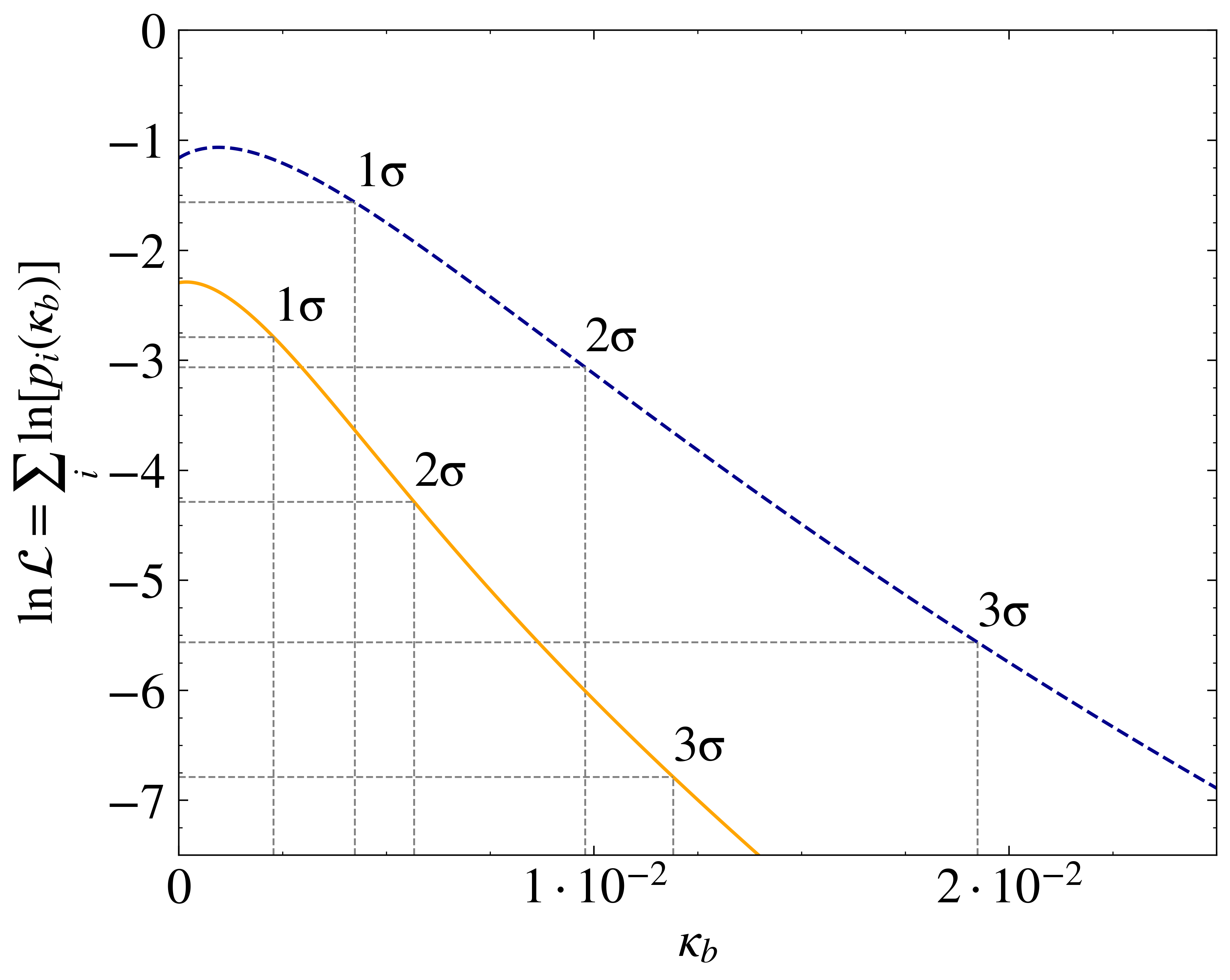}
\caption{Likelihood function $\ln \mathcal{L}$ as a function of the convergence contributed by compact objects in the dark matter, $\kappa_\text{b}$, which may be primordial black holes or other dark matter compact objects but does not include the known contribution from intracluster stars. The blue dashed line is obtained if we assume that the corrugated band is symmetric around the MaCL, while the orange line takes into account that the band is actually symmetric around the SCL.}

\label{fig:likelihood_function}
\end{figure}

The result for the likelihood function is shown in Figure \ref{fig:likelihood_function} as the solid, orange line. The optimal value
of the dark matter surface density is consistent with zero, so
the observed distribution of supermagnified star images is perfectly explained by the
estimated stellar population surface density in the measured intracluster light, and is consistent with no compact objects in the dark matter. Figure \ref{fig:likelihood_function} shows also the $(1,2,3)\sigma$ upper limits that
can be placed on any additional compact objects, if we treat the
$\ln \mathcal L$ function as a $\chi^2$ function. The $3\sigma$ upper limit,
at a confidence level of 0.3\%, is $\kappa_b < 0.012$. We warn that this upper    
limit includes only the statistical uncertainty associated with the 9 events
we used, but we have not incorporated other uncertainties due to
the modeling of the lenses, the determination of $\kappa_\star$ or the assumed mass function of the stars in the
intracluster light. This upper limit to the fraction of dark
matter in compact objects should also be slightly increased when we take into
account that $\sim$ 10\% of the total mass contributing to lensing is in the
hot, X-ray emitting gas (note that the gas usually has a density profile that
is less centrally concentrated than the dark matter in the strong lensing
region, so this fraction should be slightly smaller than the cosmic ratio of baryonic to total matter).

We show also as a blue line the result of ignoring the theoretical prediction that the corrugated band is centered on the SCL rather than the MaCL, and using equation (\ref{eq:probi}) with $s$ directly replaced by $r$, rather than equation (\ref{eq:probir}). In this case the upper limits would be increased.

As discussed at the beginning of Section \ref{sec:widths}, we have excluded from this
analysis event 6, which similarly to event 11 has a magnified image
further from the model MaCL by a factor $\sim 4$ than the predicted width $r_w$ for $\kappa_\star = 0.01 \kappa_0$, because of the model uncertainty in the
true separation from the MaCL. Had we included this event, the $3\sigma$ upper
limit on $\kappa_b$ increases only to $\kappa_b < 0.013$.

As a consequence, we can establish an upper limit to the fraction of dark matter that may be in compact objects. The average local convergence in the 9 events we have used is $\bar\kappa_0=0.691$, so the $3\sigma$ upper limit on the dark matter fraction in compact objects is $\kappa_b/\bar{\kappa}_0 \simeq 0.017$.
If we take into account the fraction of the lens surface density $\kappa_0$ that is in the hot X-ray gas, this upper limit is only slightly increased to $\simeq 2\%$.

\section{Dependence on the microlens mass distribution}
\label{sec:advanced}

Although the upper limit on the contribution of compact objects by the dark matter has been derived assuming they have the same mass as intracluster stars, the result actually extends over a very broad mass range. A remarkable result of the microlensing effect of compact masses is that the width of the corrugated network of MiCLs is simply $r_w=\kappa_\star/g$, and is independent of the mass distribution of the microlenses. The mass of the microlenses affects only the maximum magnification and frequency of the MiCL crossing events.

To see this, let us first consider the case of compact objects in the dark matter that are much more massive than stars, with Einstein radius $\theta_b\gg \theta_\star$ and convergence $\kappa_b$. These compact objects should produce their own network of MiCLs, and the intracluster stars give rise to a second level of corrugation at a smaller scale of the microcaustics, corresponding to their smaller Einstein radius. Without the intracluster stars, the peak magnification of MiCL crossings within the network is typically \citep[equations 20 and 27 of][]{VDM17}
\begin{equation}\label{eq:mpb}
\mu_{pb} \simeq \frac{1}{|1-\kappa_0|}\, \left( \frac{D_{\rm S}\theta_b}{R} \right)^{1/2}\, \kappa_b^{-3/4} ~,
\end{equation}
where $R$ is the radius of the source star. We note that this peak magnification does not actually depend on the gradient $g$, but only on $\theta_b$ and $\kappa_b$. Below a critical convergence $\kappa_{c1}=(g\theta_b)^{2/3}$, there is no corrugated network. When the intracluster stars are added, a second level of corrugation appears on the MiCLs caused by the more massive dark matter compact objects, with an external eigenvalue gradient $g_b \sim \kappa_b^{3/2}/\theta_b$ \citep[equation 26 in][]{VDM17}, if $\kappa_\star$ is greater than a new critical convergence, $\kappa_{c2}= (g_b \theta_\star)^{2/3}= \kappa_b (\theta_\star / \theta_b)^{2/3}$. The condition $\kappa_\star > \kappa_{c2}$ for the second level of corrugation to appear is the same as requiring that $\theta_\star^{1/2}/ \kappa_\star^{3/4} < \theta_b^{1/2}/\kappa_b^{3/4}$. In other words, it is the same condition as requiring that stars dominate the impact on the peak magnification, which is given by 
\begin{equation}\label{eq:mps}
\mu_{p\star} \simeq \frac{1}{ |1-\kappa_0|}\, \left( \frac{D_{\rm S}\theta_\star}{R} \right)^{1/2}\, \kappa_\star^{-3/4} ~.
\end{equation}
If $\kappa_\star < \kappa_{c2}$, then the second level of corrugation is not formed, and the peak magnification is limited by the presence of the the dark matter compact objects instead of the stars, according to equation (\ref{eq:mpb}).

In any case, we note that any restrictions that can be inferred from the maximum magnification that can be reached in MiCL crossings to allow the detection of a star, once a maximum intrinsic luminosity of the source star is assumed, are {\it in addition} to the restriction that the supermagnified images should be uniformly distributed around the MaCL within the width $r_w$ determined by the total surface density in compact objects. The first limits on $\kappa_b$ were obtained by \cite{Oguri18}, but they were based on requiring that the maximum magnification of MiCL crossings was sufficient to explain the observed flux given a maximum luminosity assumed for the source star. This limit may be subject to uncertainties arising from the possible presence of a high-magnification tail due, for example, to rare cusps on the MiCLs, and is completely independent from the limit on $\kappa_b$ we study here using the distribution of magnified images around the MaCL. In our case, the important fact is that images reaching the peak magnification of equation (\ref{eq:mps}) or (\ref{eq:mpb}), whichever is {\it lower}, should be uniformly distributed within the width $r_w$ determined by the {\it total} surface density in compact objects of masses comparable or higher than stars.

Hence, our limit implies that no more than $\sim$ 2\% of the dark matter can be in any compact objects above stellar masses, up to a maximum mass at which $\kappa_b = \kappa_{cb}=(g\theta_b)^{2/3}$. For the upper limit $\kappa_b < 0.012$, and the median value of the eigenvalue gradient for our cluster lenses $g^{-1}\simeq 16\, {\rm arcsec}$, this corresponds to $\theta_b\sim 0.05\, {\rm arcsec}$, or masses of $10^9$ to $10^{10}\msun$. Above this mass, the dark matter compact objects with $\kappa_b=0.012$ would produce isolated perturbations to the MaCL that could be individually identified around each object, which would stand out as a measurable perturbation caused by a dark object that is not associated with any luminous galaxy. Therefore, they would also be detectable in any case through their noticeable effect on cluster lensing models, above a minimum abundance that would depend on the specific observation and modeling but which is unlikely to be as high as the $\kappa_{cb}$ threshold for corrugation.

Our limit actually applies also to extended lenses that may be subcritical to lensing on their own, but become supercritical close to the cluster MaCL, such as dark matter subhalos. We note that cluster galaxies may well exceed a contribution of 2\% to the mass surface density, and their effects are in fact often visible on the shape of critical lines when they pass close to their vicinity, but this is taken into account in the lensing models with the identified luminous cluster galaxies. Similarly, if dark matter subhalos that are sufficiently dense exceed this upper limit of $\sim$ 2\% to their surface density contribution, they would have an important effect widening the region of maximally magnified images around the MaCL \citep[see][]{Dai18}, although some low-mass subhalos might remain subcritical to lensing even closer to the MaCL, and might then exceed this contribution without being easily detectable.

Next, we discuss the case of dark matter compact objects that are much less massive than stars. Exactly as in the previous case, the maximum magnification that will be reached in MiCL crossings will now be reduced to the value given in equation (\ref{eq:mpb}), where now $\theta_b < \theta_\star$, provided that $\kappa_b > \kappa_\star (\theta_b/\theta_\star)^{2/3}$. Otherwise, the peak magnifications will be the same as those caused by the intracluster stars alone. The total width of the corrugated network depends only on the total surface density of compact objects contributing to the two levels of corrugation.

However, there is in this case a different limitation to the detection of the microlensing effect: for any fixed $\kappa_b$, in the limit of very small masses of dark matter compact objects, the case of smooth dark matter must obviously be recovered. This occurs because the MiCL crossings overlap, as the separation between microcaustics is reduced below the radius of the source star. If the crossings of MiCLs can be detected down to a photometric accuracy of a fraction $f$ of the total flux, then the peak magnification of equation (\ref{eq:mpb}) must exceed $f\mu(r)$, where the mean magnification at angular separation $r$ from the MaCL is $\mu(r)\simeq |1-\kappa_0|^{-1}/(rg|\sin\alpha|)$ \citep[equation 10 in][]{VDM17}, and $\alpha$ is the angle between the MaCL and the principal axis of the vanishing eigenvalue. Applying this to the typical separation where most highly magnified images should be seen, $r\sim r_w= (\kappa_b+\kappa_\star)/g$, we obtain the condition
\begin{equation}\label{eq:tbmin}
\theta_b > \frac{R}{D_{\rm S}} \,\frac{f^2\kappa_b^{3/2}}{(\kappa_\star  +\kappa_b)^2|\sin\alpha|^2  } ~. 
\end{equation}
For the case $\kappa_b\sim \kappa_\star\sim 0.01$, and $|\sin\alpha| \simeq 1/2$ and $f\sim 1/2$, we find that $D_{\rm S}\theta_b/R \gtrsim 2.5$, and for a typical stellar size and source angular diameter distance $R/D_{\rm S}\sim 10^{-15}$, we find $\theta_b \gtrsim 5\times 10^{-10}\, {\rm arcsec}$, implying that masses down to $10^{-6}\msun$ are still ruled out to contribute a surface density greater than $\kappa_b \gtrsim 0.012$ because any highly magnified, variable images would be observed over a width larger than observed around the MaCL.

Note that for the microlensing mass that produces a typical magnification peak equal to the average magnification (i.e., when equation (\ref{eq:tbmin}) is an equality with $f=1$), this peak magnification is
\begin{equation}
 \mu_{pb}= \frac{1}{|(1-\kappa_0)\sin\alpha|(\kappa_\star+\kappa_b) } ~, 
\end{equation}
as expected at the separation $\sim r_w$ from the MaCL. For the minimum mass corresponding to the Einstein radius $\theta_b=(R/D_s)\kappa_b^{3/2}(\kappa_\star+\kappa_b)^{-2}|\sin\alpha|^{-2}$, the microcaustic peaks are overlapping over the whole corrugated network; in other words, the mean separation between neighboring microcaustics is equal to the source radius.

\section{Conclusions}
\label{sec:conclusions}

The result presented in this paper implies that the observation of only 9 supermagnified stars in
lensing clusters of galaxies places an upper limit of 2\% on the fraction of dark
matter in compact objects.
This is already
better than upper limits obtained from several microlensing surveys
that have monitored millions of stars in the Magellanic Clouds \citep{MACHO01,EROS07,OGLE3-11b}, except for the recent OGLE result which achieves an even stronger upper limit \citep{Mroz24}.
Models of primordial black holes making up the dark matter are now ruled out for masses $m_b \gtrsim 10^{-6}\msun$ for fractions higher than $\sim$ 2\% at the 99\% confidence level. Interestingly, this limit applies to the dark matter that is present in the most massive clusters of galaxies in the Universe, and also to the intergalactic dark matter that must be present along the line of sight to the source that traverses the lensing cluster, and contributes to the total lensing convergence.

As the observations of variability in lensing clusters improve in the future with further observations of JWST and other telescopes, we can look forward to an improvement in these limits, and also to new constraints on the mass function of the microlenses caused by intracluster stars, or perhaps other unknown objects. Other models such as the QCD axion, which predicts the presence of axion minihalos that can also have lensing effects despite their extended mass distribution \citep{DM20}, will likely be constrained as well by monitoring campaigns of the supermagnified stars discovered by \cite{KDR18}. 

\section{Acknowledgments}

This work has used gravitational lensing models produced by PIs Bradač, Natarajan and Kneib (CATS), Merten and Zitrin, Sharon, Williams, Keeton, Bernstein and Diego, and the GLAFIC group. This lens modeling was partially funded by the HST Frontier Fields program conducted by STScI. STScI is operated by the Association of Universities for Research in Astronomy, Inc. under NASA contract NAS 5-26555. The lens models were obtained from the Mikulski Archive for Space Telescopes (MAST).
This work is also based on observations taken by the RELICS Treasury Program (GO 14096) with the NASA/ESA HST, which is operated by the Association of Universities for Research in Astronomy, Inc., under NASA contract NAS5-26555.
This work was supported in part by Spanish grants PID2019-108122GB-C32 and PID2022-137268NB-C52.

\section{Data Availability}
The data and code used for the work presented in this article are available at \url{https://doi.org/10.6084/m9.figshare.27635736.v1}.

\newpage
\nocite{*}

\bibliographystyle{mnras}
\bibliography{references}

\begin{thebibliography}{}
\makeatletter
\relax
\def\mn@urlcharsother{\let\do\@makeother \do\$\do\&\do\#\do\^\do\_\do\%\do\~}
\def\mn@doi{\begingroup\mn@urlcharsother \@ifnextchar [ {\mn@doi@} {\mn@doi@[]}}
\def\mn@doi@[#1]#2{\def\@tempa{#1}\ifx\@tempa\@empty \href {http://dx.doi.org/#2} {doi:#2}\else \href {http://dx.doi.org/#2} {#1}\fi \endgroup}
\def\mn@eprint#1#2{\mn@eprint@#1:#2::\@nil}
\def\mn@eprint@arXiv#1{\href {http://arxiv.org/abs/#1} {{\tt arXiv:#1}}}
\def\mn@eprint@dblp#1{\href {http://dblp.uni-trier.de/rec/bibtex/#1.xml} {dblp:#1}}
\def\mn@eprint@#1:#2:#3:#4\@nil{\def\@tempa {#1}\def\@tempb {#2}\def\@tempc {#3}\ifx \@tempc \@empty \let \@tempc \@tempb \let \@tempb \@tempa \fi \ifx \@tempb \@empty \def\@tempb {arXiv}\fi \@ifundefined {mn@eprint@\@tempb}{\@tempb:\@tempc}{\expandafter \expandafter \csname mn@eprint@\@tempb\endcsname \expandafter{\@tempc}}}

\bibitem[\protect\citeauthoryear{{Alam} et~al.,}{{Alam} et~al.}{2021}]{EBOSS21}
{Alam} S.,  et~al., 2021, \mn@doi [\prd] {10.1103/PhysRevD.103.083533}, \href {https://ui.adsabs.harvard.edu/abs/2021PhRvD.103h3533A} {103, 083533}

\bibitem[\protect\citeauthoryear{{Alcock} et~al.,}{{Alcock} et~al.}{1998}]{MACHO98}
{Alcock} C.,  et~al., 1998, \mn@doi [\apjl] {10.1086/311355}, \href {https://ui.adsabs.harvard.edu/abs/1998ApJ...499L...9A} {499, L9}

\bibitem[\protect\citeauthoryear{{Alcock} et~al.,}{{Alcock} et~al.}{2001}]{MACHO01}
{Alcock} C.,  et~al., 2001, \mn@doi [\apjl] {10.1086/319636}, \href {https://ui.adsabs.harvard.edu/abs/2001ApJ...550L.169A} {550, L169}

\bibitem[\protect\citeauthoryear{{Ammons}, {Wong}, {Zabludoff}  \& {Keeton}}{{Ammons} et~al.}{2014}]{Ammons14LS1Keeton4}
{Ammons} S.~M.,  {Wong} K.~C.,  {Zabludoff} A.~I.,   {Keeton} C.~R.,  2014, \mn@doi [\apj] {10.1088/0004-637X/781/1/2}, \href {https://ui.adsabs.harvard.edu/abs/2014ApJ...781....2A} {781, 2}

\bibitem[\protect\citeauthoryear{{Brada{\v{c}}}, {Schneider}, {Lombardi}  \& {Erben}}{{Brada{\v{c}}} et~al.}{2005}]{Bradac05LS2Bradac3}
{Brada{\v{c}}} M.,  {Schneider} P.,  {Lombardi} M.,   {Erben} T.,  2005, \mn@doi [\aap] {10.1051/0004-6361:20042233}, \href {https://ui.adsabs.harvard.edu/abs/2005A&A...437...39B} {437, 39}

\bibitem[\protect\citeauthoryear{{Brada{\v{c}}} et~al.,}{{Brada{\v{c}}} et~al.}{2009}]{Bradac09LS2Bradac3}
{Brada{\v{c}}} M.,  et~al., 2009, \mn@doi [\apj] {10.1088/0004-637X/706/2/1201}, \href {https://ui.adsabs.harvard.edu/abs/2009ApJ...706.1201B} {706, 1201}

\bibitem[\protect\citeauthoryear{{Caminha} et~al.,}{{Caminha} et~al.}{2017}]{Caminha17LS4Caminha4}
{Caminha} G.~B.,  et~al., 2017, \mn@doi [\aap] {10.1051/0004-6361/201629297}, \href {https://ui.adsabs.harvard.edu/abs/2017A&A...600A..90C} {600, A90}

\bibitem[\protect\citeauthoryear{{Carr}, {Kohri}, {Sendouda}  \& {Yokoyama}}{{Carr} et~al.}{2021}]{Carr21}
{Carr} B.,  {Kohri} K.,  {Sendouda} Y.,   {Yokoyama} J.,  2021, \mn@doi [Reports on Progress in Physics] {10.1088/1361-6633/ac1e31}, \href {https://ui.adsabs.harvard.edu/abs/2021RPPh...84k6902C} {84, 116902}

\bibitem[\protect\citeauthoryear{{Chang} \& {Refsdal}}{{Chang} \& {Refsdal}}{1979}]{Chang79}
{Chang} K.,  {Refsdal} S.,  1979, \mn@doi [\nat] {10.1038/282561a0}, \href {https://ui.adsabs.harvard.edu/abs/1979Natur.282..561C} {282, 561}

\bibitem[\protect\citeauthoryear{{Chen} et~al.,}{{Chen} et~al.}{2022}]{CKT22}
{Chen} W.,  et~al., 2022, \mn@doi [\apjl] {10.3847/2041-8213/ac9585}, \href {https://ui.adsabs.harvard.edu/abs/2022ApJ...940L..54C} {940, L54}

\bibitem[\protect\citeauthoryear{{Coe} et~al.,}{{Coe} et~al.}{2019}]{Coe19RELICS}
{Coe} D.,  et~al., 2019, \mn@doi [\apj] {10.3847/1538-4357/ab412b}, \href {https://ui.adsabs.harvard.edu/abs/2019ApJ...884...85C} {884, 85}

\bibitem[\protect\citeauthoryear{{Dai} \& {Miralda-Escud{\'e}}}{{Dai} \& {Miralda-Escud{\'e}}}{2020}]{DM20}
{Dai} L.,  {Miralda-Escud{\'e}} J.,  2020, \mn@doi [\aj] {10.3847/1538-3881/ab5e83}, \href {https://ui.adsabs.harvard.edu/abs/2020AJ....159...49D} {159, 49}

\bibitem[\protect\citeauthoryear{{Dai}, {Venumadhav}, {Kaurov}  \& {Miralda-Escud\'e}}{{Dai} et~al.}{2018}]{Dai18}
{Dai} L.,  {Venumadhav} T.,  {Kaurov} A.~A.,   {Miralda-Escud\'e} J.,  2018, \mn@doi [\apj] {10.3847/1538-4357/aae478}, \href {https://ui.adsabs.harvard.edu/abs/2018ApJ...867...24D} {867, 24}

\bibitem[\protect\citeauthoryear{{Diego}, {Pascale}, {Kavanagh}, {Kelly}, {Dai}, {Frye}  \& {Broadhurst}}{{Diego} et~al.}{2022}]{DPK22}
{Diego} J.~M.,  {Pascale} M.,  {Kavanagh} B.~J.,  {Kelly} P.,  {Dai} L.,  {Frye} B.,   {Broadhurst} T.,  2022, \mn@doi [\aap] {10.1051/0004-6361/202243605}, \href {https://ui.adsabs.harvard.edu/abs/2022A&A...665A.134D} {665, A134}

\bibitem[\protect\citeauthoryear{{Diego} et~al.,}{{Diego} et~al.}{2023a}]{DSY23}
{Diego} J.~M.,  et~al., 2023a, \mn@doi [arXiv e-prints] {10.48550/arXiv.2307.10363}, \href {https://ui.adsabs.harvard.edu/abs/2023arXiv230710363D} {p. arXiv:2307.10363}

\bibitem[\protect\citeauthoryear{{Diego} et~al.,}{{Diego} et~al.}{2023b}]{DMA23}
{Diego} J.~M.,  et~al., 2023b, \mn@doi [\aap] {10.1051/0004-6361/202245238}, \href {https://ui.adsabs.harvard.edu/abs/2023A&A...672A...3D} {672, A3}

\bibitem[\protect\citeauthoryear{{Einasto}, {Kaasik}  \& {Saar}}{{Einasto} et~al.}{1974}]{EKS74}
{Einasto} J.,  {Kaasik} A.,   {Saar} E.,  1974, \mn@doi [\nat] {10.1038/250309a0}, \href {https://ui.adsabs.harvard.edu/abs/1974Natur.250..309E} {250, 309}

\bibitem[\protect\citeauthoryear{{Griest}, {Cieplak}  \& {Lehner}}{{Griest} et~al.}{2014}]{Griest14}
{Griest} K.,  {Cieplak} A.~M.,   {Lehner} M.~J.,  2014, \mn@doi [\apj] {10.1088/0004-637X/786/2/158}, \href {https://ui.adsabs.harvard.edu/abs/2014ApJ...786..158G} {786, 158}

\bibitem[\protect\citeauthoryear{{Grohs} \& {Fuller}}{{Grohs} \& {Fuller}}{2022}]{GF22}
{Grohs} E.,  {Fuller} G.~M.,  2022, in , Handbook of Nuclear Physics.
p.~127, \mn@doi{10.1007/978-981-15-8818-1_127-1}

\bibitem[\protect\citeauthoryear{{Hawking}}{{Hawking}}{1971}]{Hawking71}
{Hawking} S.,  1971, \mn@doi [\mnras] {10.1093/mnras/152.1.75}, \href {https://ui.adsabs.harvard.edu/abs/1971MNRAS.152...75H} {152, 75}

\bibitem[\protect\citeauthoryear{{Hoag} et~al.,}{{Hoag} et~al.}{2016}]{Hoag16LS2Bradac3}
{Hoag} A.,  et~al., 2016, \mn@doi [\apj] {10.3847/0004-637X/831/2/182}, \href {https://ui.adsabs.harvard.edu/abs/2016ApJ...831..182H} {831, 182}

\bibitem[\protect\citeauthoryear{{Johnson}, {Sharon}, {Bayliss}, {Gladders}, {Coe}  \& {Ebeling}}{{Johnson} et~al.}{2014}]{Johnson14LS4Sharon4.2}
{Johnson} T.~L.,  {Sharon} K.,  {Bayliss} M.~B.,  {Gladders} M.~D.,  {Coe} D.,   {Ebeling} H.,  2014, \mn@doi [\apj] {10.1088/0004-637X/797/1/48}, \href {https://ui.adsabs.harvard.edu/abs/2014ApJ...797...48J} {797, 48}

\bibitem[\protect\citeauthoryear{{Jullo}, {Kneib}, {Limousin}, {El{\'\i}asd{\'o}ttir}, {Marshall}  \& {Verdugo}}{{Jullo} et~al.}{2007}]{Jullo07LS4Sharon4.2}
{Jullo} E.,  {Kneib} J.~P.,  {Limousin} M.,  {El{\'\i}asd{\'o}ttir} {\'A}.,  {Marshall} P.~J.,   {Verdugo} T.,  2007, \mn@doi [New Journal of Physics] {10.1088/1367-2630/9/12/447}, \href {https://ui.adsabs.harvard.edu/abs/2007NJPh....9..447J} {9, 447}

\bibitem[\protect\citeauthoryear{{Kaurov}, {Dai}, {Venumadhav}, {Miralda-Escud{\'e}}  \& {Frye}}{{Kaurov} et~al.}{2019}]{KDV19}
{Kaurov} A.~A.,  {Dai} L.,  {Venumadhav} T.,  {Miralda-Escud{\'e}} J.,   {Frye} B.,  2019, \mn@doi [\apj] {10.3847/1538-4357/ab2888}, \href {https://ui.adsabs.harvard.edu/abs/2019ApJ...880...58K} {880, 58}

\bibitem[\protect\citeauthoryear{{Kawamata}, {Oguri}, {Ishigaki}, {Shimasaku}  \& {Ouchi}}{{Kawamata} et~al.}{2016}]{Kawamata16LS1GLAFIC3}
{Kawamata} R.,  {Oguri} M.,  {Ishigaki} M.,  {Shimasaku} K.,   {Ouchi} M.,  2016, \mn@doi [\apj] {10.3847/0004-637X/819/2/114}, \href {https://ui.adsabs.harvard.edu/abs/2016ApJ...819..114K} {819, 114}

\bibitem[\protect\citeauthoryear{{Kawamata}, {Ishigaki}, {Shimasaku}, {Oguri}, {Ouchi}  \& {Tanigawa}}{{Kawamata} et~al.}{2018}]{Kawamata18LS4Glafic4}
{Kawamata} R.,  {Ishigaki} M.,  {Shimasaku} K.,  {Oguri} M.,  {Ouchi} M.,   {Tanigawa} S.,  2018, \mn@doi [\apj] {10.3847/1538-4357/aaa6cf}, \href {https://ui.adsabs.harvard.edu/abs/2018ApJ...855....4K} {855, 4}

\bibitem[\protect\citeauthoryear{{Keeton}}{{Keeton}}{2010}]{Keeton10LS1Keeton4}
{Keeton} C.~R.,  2010, \mn@doi [General Relativity and Gravitation] {10.1007/s10714-010-1041-1}, \href {https://ui.adsabs.harvard.edu/abs/2010GReGr..42.2151K} {42, 2151}

\bibitem[\protect\citeauthoryear{{Kelly} et~al.,}{{Kelly} et~al.}{2018}]{KDR18}
{Kelly} P.~L.,  et~al., 2018, \mn@doi [Nature Astronomy] {10.1038/s41550-018-0430-3}, \href {https://ui.adsabs.harvard.edu/abs/2018NatAs...2..334K} {2, 334}

\bibitem[\protect\citeauthoryear{{Liesenborgs}, {De Rijcke}  \& {Dejonghe}}{{Liesenborgs} et~al.}{2006}]{Liesenborgs06LS2Williams4}
{Liesenborgs} J.,  {De Rijcke} S.,   {Dejonghe} H.,  2006, \mn@doi [\mnras] {10.1111/j.1365-2966.2006.10040.x}, \href {https://ui.adsabs.harvard.edu/abs/2006MNRAS.367.1209L} {367, 1209}

\bibitem[\protect\citeauthoryear{{Lotz} et~al.,}{{Lotz} et~al.}{2017}]{LKC17-HFF}
{Lotz} J.~M.,  et~al., 2017, \mn@doi [\apj] {10.3847/1538-4357/837/1/97}, \href {https://ui.adsabs.harvard.edu/abs/2017ApJ...837...97L} {837, 97}

\bibitem[\protect\citeauthoryear{{Madhavacheril} et~al.,}{{Madhavacheril} et~al.}{2023}]{Mad23}
{Madhavacheril} M.~S.,  et~al., 2023, \mn@doi [arXiv e-prints] {10.48550/arXiv.2304.05203}, \href {https://ui.adsabs.harvard.edu/abs/2023arXiv230405203M} {p. arXiv:2304.05203}

\bibitem[\protect\citeauthoryear{{McCully}, {Keeton}, {Wong}  \& {Zabludoff}}{{McCully} et~al.}{2014}]{McCully14LS1Keeton4}
{McCully} C.,  {Keeton} C.~R.,  {Wong} K.~C.,   {Zabludoff} A.~I.,  2014, \mn@doi [\mnras] {10.1093/mnras/stu1316}, \href {https://ui.adsabs.harvard.edu/abs/2014MNRAS.443.3631M} {443, 3631}

\bibitem[\protect\citeauthoryear{{Meena} et~al.,}{{Meena} et~al.}{2023a}]{MCZ23}
{Meena} A.~K.,  et~al., 2023a, \mn@doi [\mnras] {10.1093/mnras/stad869}, \href {https://ui.adsabs.harvard.edu/abs/2023MNRAS.521.5224M} {521, 5224}

\bibitem[\protect\citeauthoryear{{Meena} et~al.,}{{Meena} et~al.}{2023b}]{MZJ23}
{Meena} A.~K.,  et~al., 2023b, \mn@doi [\apjl] {10.3847/2041-8213/acb645}, \href {https://ui.adsabs.harvard.edu/abs/2023ApJ...944L...6M} {944, L6}

\bibitem[\protect\citeauthoryear{{Miralda-Escud\'e}}{{Miralda-Escud\'e}}{1991}]{Miralda91}
{Miralda-Escud\'e} J.,  1991, \mn@doi [\apj] {10.1086/170486}, \href {https://ui.adsabs.harvard.edu/abs/1991ApJ...379...94M} {379, 94}

\bibitem[\protect\citeauthoryear{{Mr{\'o}z} et~al.,}{{Mr{\'o}z} et~al.}{2024a}]{Mroz24}
{Mr{\'o}z} P.,  et~al., 2024a, \mn@doi [\apjs] {10.3847/1538-4365/ad452e}, \href {https://ui.adsabs.harvard.edu/abs/2024ApJS..273....4M} {273, 4}

\bibitem[\protect\citeauthoryear{{Mr{\'o}z} et~al.,}{{Mr{\'o}z} et~al.}{2024b}]{MrozN24}
{Mr{\'o}z} P.,  et~al., 2024b, \mn@doi [\nat] {10.1038/s41586-024-07704-6}, \href {https://ui.adsabs.harvard.edu/abs/2024Natur.632..749M} {632, 749}

\bibitem[\protect\citeauthoryear{{Oguri}}{{Oguri}}{2010}]{Oguri10LS1GLAFIC3}
{Oguri} M.,  2010, \mn@doi [\pasj] {10.1093/pasj/62.4.1017}, \href {https://ui.adsabs.harvard.edu/abs/2010PASJ...62.1017O} {62, 1017}

\bibitem[\protect\citeauthoryear{{Oguri}, {Diego}, {Kaiser}, {Kelly}  \& {Broadhurst}}{{Oguri} et~al.}{2018}]{Oguri18}
{Oguri} M.,  {Diego} J.~M.,  {Kaiser} N.,  {Kelly} P.~L.,   {Broadhurst} T.,  2018, \mn@doi [\prd] {10.1103/PhysRevD.97.023518}, \href {https://ui.adsabs.harvard.edu/abs/2018PhRvD..97b3518O} {97, 023518}

\bibitem[\protect\citeauthoryear{{Ostriker}, {Peebles}  \& {Yahil}}{{Ostriker} et~al.}{1974}]{OPY74}
{Ostriker} J.~P.,  {Peebles} P.~J.~E.,   {Yahil} A.,  1974, \mn@doi [\apjl] {10.1086/181617}, \href {https://ui.adsabs.harvard.edu/abs/1974ApJ...193L...1O} {193, L1}

\bibitem[\protect\citeauthoryear{{Peebles}}{{Peebles}}{1982}]{P82}
{Peebles} P.~J.~E.,  1982, \mn@doi [\apjl] {10.1086/183911}, \href {https://ui.adsabs.harvard.edu/abs/1982ApJ...263L...1P} {263, L1}

\bibitem[\protect\citeauthoryear{{Planck Collaboration} et~al.,}{{Planck Collaboration} et~al.}{2020}]{Planck20}
{Planck Collaboration} et~al., 2020, \mn@doi [\aap] {10.1051/0004-6361/201833910}, \href {https://ui.adsabs.harvard.edu/abs/2020A&A...641A...6P} {641, A6}

\bibitem[\protect\citeauthoryear{{Postman} et~al.,}{{Postman} et~al.}{2012}]{Postman12CLASH}
{Postman} M.,  et~al., 2012, \mn@doi [\apjs] {10.1088/0067-0049/199/2/25}, \href {https://ui.adsabs.harvard.edu/abs/2012ApJS..199...25P} {199, 25}

\bibitem[\protect\citeauthoryear{{Richard} et~al.,}{{Richard} et~al.}{2014}]{Richard14LS1Cats4.1}
{Richard} J.,  et~al., 2014, \mn@doi [\mnras] {10.1093/mnras/stu1395}, \href {https://ui.adsabs.harvard.edu/abs/2014MNRAS.444..268R} {444, 268}

\bibitem[\protect\citeauthoryear{{Rodney} et~al.,}{{Rodney} et~al.}{2018}]{RBB18}
{Rodney} S.~A.,  et~al., 2018, \mn@doi [Nature Astronomy] {10.1038/s41550-018-0405-4}, \href {https://ui.adsabs.harvard.edu/abs/2018NatAs...2..324R} {2, 324}

\bibitem[\protect\citeauthoryear{{Rubin}, {Ford}  \& {Thonnard}}{{Rubin} et~al.}{1978}]{Rubin78}
{Rubin} V.~C.,  {Ford} W.~K. J.,   {Thonnard} N.,  1978, \mn@doi [\apjl] {10.1086/182804}, \href {https://ui.adsabs.harvard.edu/abs/1978ApJ...225L.107R} {225, L107}

\bibitem[\protect\citeauthoryear{{Sebesta}, {Williams}, {Mohammed}, {Saha}  \& {Liesenborgs}}{{Sebesta} et~al.}{2016}]{Sebesta16LS2Williams4}
{Sebesta} K.,  {Williams} L. L.~R.,  {Mohammed} I.,  {Saha} P.,   {Liesenborgs} J.,  2016, \mn@doi [\mnras] {10.1093/mnras/stw1433}, \href {https://ui.adsabs.harvard.edu/abs/2016MNRAS.461.2126S} {461, 2126}

\bibitem[\protect\citeauthoryear{{Sebesta}, {Williams}, {Liesenborgs}, {Medezinski}  \& {Okabe}}{{Sebesta} et~al.}{2019}]{Sebesta19LS11Williams1}
{Sebesta} K.,  {Williams} L. L.~R.,  {Liesenborgs} J.,  {Medezinski} E.,   {Okabe} N.,  2019, \mn@doi [\mnras] {10.1093/mnras/stz1950}, \href {https://ui.adsabs.harvard.edu/abs/2019MNRAS.488.3251S} {488, 3251}

\bibitem[\protect\citeauthoryear{{Sendra}, {Diego}, {Broadhurst}  \& {Lazkoz}}{{Sendra} et~al.}{2014}]{Sendra14LS5WSLAP1}
{Sendra} I.,  {Diego} J.~M.,  {Broadhurst} T.,   {Lazkoz} R.,  2014, \mn@doi [\mnras] {10.1093/mnras/stt2076}, \href {https://ui.adsabs.harvard.edu/abs/2014MNRAS.437.2642S} {437, 2642}

\bibitem[\protect\citeauthoryear{{Smyth}, {Profumo}, {English}, {Jeltema}, {McKinnon}  \& {Guhathakurta}}{{Smyth} et~al.}{2020}]{Smyth20}
{Smyth} N.,  {Profumo} S.,  {English} S.,  {Jeltema} T.,  {McKinnon} K.,   {Guhathakurta} P.,  2020, \mn@doi [\prd] {10.1103/PhysRevD.101.063005}, \href {https://ui.adsabs.harvard.edu/abs/2020PhRvD.101f3005S} {101, 063005}

\bibitem[\protect\citeauthoryear{{Tisserand} et~al.,}{{Tisserand} et~al.}{2007}]{EROS07}
{Tisserand} P.,  et~al., 2007, \mn@doi [\aap] {10.1051/0004-6361:20066017}, \href {https://ui.adsabs.harvard.edu/abs/2007A&A...469..387T} {469, 387}

\bibitem[\protect\citeauthoryear{{Tristram} et~al.,}{{Tristram} et~al.}{2024}]{T24}
{Tristram} M.,  et~al., 2024, \mn@doi [\aap] {10.1051/0004-6361/202348015}, \href {https://ui.adsabs.harvard.edu/abs/2024A&A...682A..37T} {682, A37}

\bibitem[\protect\citeauthoryear{{Umetsu} et~al.,}{{Umetsu} et~al.}{2018}]{Umetsu18}
{Umetsu} K.,  et~al., 2018, \mn@doi [\apj] {10.3847/1538-4357/aac3d9}, \href {https://ui.adsabs.harvard.edu/abs/2018ApJ...860..104U} {860, 104}

\bibitem[\protect\citeauthoryear{{Venumadhav}, {Dai}  \& {Miralda-Escud{\'e}}}{{Venumadhav} et~al.}{2017}]{VDM17}
{Venumadhav} T.,  {Dai} L.,   {Miralda-Escud{\'e}} J.,  2017, \mn@doi [\apj] {10.3847/1538-4357/aa9575}, \href {https://ui.adsabs.harvard.edu/abs/2017ApJ...850...49V} {850, 49}

\bibitem[\protect\citeauthoryear{{Welch} et~al.,}{{Welch} et~al.}{2022}]{WCD22}
{Welch} B.,  et~al., 2022, \mn@doi [\nat] {10.1038/s41586-022-04449-y}, \href {https://ui.adsabs.harvard.edu/abs/2022Natur.603..815W} {603, 815}

\bibitem[\protect\citeauthoryear{{Wyrzykowski} et~al.,}{{Wyrzykowski} et~al.}{2011a}]{OGLE3-11a}
{Wyrzykowski} {\L}.,  et~al., 2011a, \mn@doi [\mnras] {10.1111/j.1365-2966.2010.18150.x}, \href {https://ui.adsabs.harvard.edu/abs/2011MNRAS.413..493W} {413, 493}

\bibitem[\protect\citeauthoryear{{Wyrzykowski} et~al.,}{{Wyrzykowski} et~al.}{2011b}]{OGLE3-11b}
{Wyrzykowski} L.,  et~al., 2011b, \mn@doi [\mnras] {10.1111/j.1365-2966.2011.19243.x}, \href {https://ui.adsabs.harvard.edu/abs/2011MNRAS.416.2949W} {416, 2949}

\bibitem[\protect\citeauthoryear{{Zitrin} et~al.,}{{Zitrin} et~al.}{2009}]{Zitrin09LS1Zltm1}
{Zitrin} A.,  et~al., 2009, \mn@doi [\mnras] {10.1111/j.1365-2966.2009.14899.x}, \href {https://ui.adsabs.harvard.edu/abs/2009MNRAS.396.1985Z} {396, 1985}

\bibitem[\protect\citeauthoryear{{Zitrin} et~al.,}{{Zitrin} et~al.}{2013}]{Zitrin13LS1Zltm1}
{Zitrin} A.,  et~al., 2013, \mn@doi [\apjl] {10.1088/2041-8205/762/2/L30}, \href {https://ui.adsabs.harvard.edu/abs/2013ApJ...762L..30Z} {762, L30}

\bibitem[\protect\citeauthoryear{{Zitrin} et~al.,}{{Zitrin} et~al.}{2015}]{Zitrin15LS5Zltm1}
{Zitrin} A.,  et~al., 2015, \mn@doi [\apj] {10.1088/0004-637X/801/1/44}, \href {https://ui.adsabs.harvard.edu/abs/2015ApJ...801...44Z} {801, 44}

\bibitem[\protect\citeauthoryear{{Zwicky}}{{Zwicky}}{1933}]{Zwicky33}
{Zwicky} F.,  1933, Helvetica Physica Acta, \href {https://ui.adsabs.harvard.edu/abs/1933AcHPh...6..110Z} {6, 110}

\makeatother
\end{thebibliography}

\bsp	
\label{lastpage}
\end{document}